\documentclass[manuscript]{aastex}
\usepackage{epsfig}
\usepackage{graphicx}
\usepackage{dcolumn}
\usepackage{bm}
\usepackage{booktabs}
\usepackage{amsmath}
\usepackage{lineno}
\usepackage{cite}
\usepackage{lineno}

\usepackage[
top    = 2.75cm,
bottom = 2.50cm,
left   = 1.50cm,
right  = 1.50cm]
{geometry}

\shorttitle{New thermonuclear rate of $^7$Li(d,n)2$^4$He relevant to the Cosmological Lithium Problem}
\shortauthors{S.Q. Hou et al.}

\usepackage{xcolor}
\begin{document}


\title{New Thermonuclear Rate of $^7$Li(d,n)2$^4$He relevant to the Cosmological Lithium Problem}


\author{S. Q. Hou\altaffilmark{1,2,3}, T. Kajino\altaffilmark{4,5,6}, T. C. L. Trueman\altaffilmark{7,8,3}, M. Pignatari\altaffilmark{7,8,3,9}, Y. D. Luo\altaffilmark{4,5}, C. A. Bertulani\altaffilmark{10}}


\affil{\altaffilmark{1}Key Laboratory of High Precision Nuclear Spectroscopy, Institute of Modern Physics, Chinese Academy of Sciences, Lanzhou 730000, China; sqhou@impcas.ac.cn}
\affil{\altaffilmark{2}School of Nuclear Science and Technology, University of Chinese Academy of Sciences, Beijing 100049, China}
\affil{\altaffilmark{3} NuGrid Collaboration, http://www.nugridstars.org}


\affil{\altaffilmark{4}National Astronomical Observatory of Japan, 2-21-1 Osawa, Mitaka, Tokyo 181-8588, Japan; kajino@nao.ac.jp}
\affil{\altaffilmark{5}Graduate School of Science, University of Tokyo, 7-3-1 Hongo, Bunkyo-ku, Tokyo 113-0033, Japan}
\affil{\altaffilmark{6} School of Physics, and International Research Center for Big-Bang Cosmology and Element Genesis, Beihang University 37, Xueyuan Road, Haidian-qu, Beijing 100083, People's Republic of China}

\affil{\altaffilmark{7} E. A. Milne Centre for Astrophysics, Department of Physics and Mathematics, University of Hull, Kingston upon Hull HU6 7RX, United Kingdom}
\affil{\altaffilmark{8} Konkoly Observatory, Research Centre for Astronomy and Earth Sciences, Hungarian Academy of Sciences, H-1121 Budapest, Hungary}
\affil{\altaffilmark{9} Joint Institute for Nuclear Astrophysics, Center for the Evolution of the Elements, Michigan State University, East Lansing, Michigan 48824, USA}
\affil{\altaffilmark{10} Department of Physics and Astromomy, Texas A\&M University-Commerce, Commerce, Texas 75429, USA }
\affil{\altaffilmark{11} }

\email{sqhou@impcas.ac.cn}

\begin{abstract}
Accurate $^7$Li(d,n)2$^4$He thermonuclear reaction rates are crucial for precise prediction of the primordial abundances of Lithium and Beryllium and to probe the mysteries beyond fundamental physics and the standard cosmological model. However, uncertainties still exist in current reaction rates of $^7$Li(d,n)2$^4$He widely used in Big Bang Nucleosynthesis (BBN) simulations. In this work, we reevaluate the $^7$Li(d,n)2$^4$He reaction rate using the latest data on the three near-threshold $^9$Be excited states from experimental measurements. We present for the first time uncertainties that are directly constrained by experiments.
Additionally, we take into account for the first time the contribution from the subthreshold resonance at 16.671 MeV of $^9$Be. We obtain a $^7$Li(d,n)2$^4$He rate that is overall smaller than the previous estimation by about a factor of 60 at the typical temperature of the onset of primordial nucleosynthesis.
We implemented our new rate in BBN nucleosynthesis calculations, and we show that the new rates have a very limited impact on the final light element abundances in uniform density models. Typical abundance variations are in the order of 0.002\%. For non-uniform density BBN models, the predicted $^7$Li production can be increased by 10\% and the primordial production of light nuclides with mass number A$\textgreater$7 can be increased by about 40\%. Our results confirm that the cosmological lithium problem remains a long-standing unresolved puzzle from the standpoint of nuclear physics.

\end{abstract}


\section{Introduction}
Regarded as a key pillar of modern cosmology, Big Bang nucleosynthesis (BBN) describes the conditions in which nuclear reactions built the first complex nuclei as the universe expanded and cooled down from an incredibly dense and hot primordial fireball. About half an hour after the start of the Big Bang, the entire process of primordial nucleosynthesis ended, leaving behind as main relics $^2$H, $^3$He, $^4$He, $^7$Li. In the standard BBN model, the prediction of primordial abundances depend on only one free parameter: the baryon-to-photo ratio $\eta$, which has been determined quite accurately from observations of the anisotropies of the Cosmic Microwave Backgroud (CMB)~\citep{planc}. Thus, provided that no uncertainties exist in the relevant reaction rates, the BBN-predicted light nuclide abundances should be reliable. Current BBN predictions for abundances of D, $^3$He, $^4$He are consistent with values inferred from astronomical observations. However, only the $^7$Li abundance is over-predicted by about a factor of three~\citep{cyb03,coc04,asp06,sbo10}. This is called the "cosmological lithium problem".

Over the past decade, many attempts to address this issue have been carried out, such as from the perspective of conventional nuclear physics and even exotic physics beyond the standard BBN framework~\citep{ang05,
cyb08,boy10,pos10,fie11,wan11,kir11,bro12,cyb12,tor12,coc12,kan12,ham13,coc13,piz14,kus14,coc14,yam14,
hou15,fam16,cyb16,hou17,dam18,mic18,rij19,luo19,cla20}. However, despite the fact some solutions using exotic-physics have succeeded in resolving this issue, it appears there is still no universally accepted solution in the academic community since validations of these mysterious exotic-physics are beyond the capabilities of current science. Conversely, it seems more worthwhile to exclude any potential possibility of resolving the $^7$Li discrepancy from the perspective of nuclear physics  and many have argued this since the problem was first put forward~\citep{kir11,bro12,ham13,hou15,mic18}. 
Meanwhile, enormous efforts have been made to refine the reaction rates of key BBN reactions in the past twenty years~\citep{smi93,ser04,des04,cy08b,nif11,tum14,piz14,hou15,ili16,liv17,bar16,kaw17,dam18,liv19,rij19,mos20}, but the probability of solving or alleviating the $^7$Li problem by improving our knowledge of relevant nuclear reaction rates still cannot be eliminated. 
Recent experiments for key nuclear reactions like $^7$Be(n,p)$^7$Li and $^7$Be(d,p)2$^4$He allow for a reduction of the $^7$Li production by about 12\%~\citep{dam18,rij19} compared to previous calculations. At present, nuclear uncertainties cannot rule out that some of the reactions destroying $^7$Li are indeed more efficient than those currently used~\citep{cha11, boy10}.

Despite the fact it is an important $^7$Li destruction reaction,  before 2018 the $^7$Li(d,n)2$^4$He reaction could not been well studied due to limited information of energy levels close to the threshold in $^9$Be. With mounting experimental results concerning the properties of relevant excited states emerging recently, it is the right time to reinvestigate the $^7$Li(d,n)2$^4$He reaction rate. The necessity of carrying out this work can be summarized by: Firstly, the maximum reduction of uncertainties of $^7$Li(d,n)2$^4$He reaction rates can remove the most significant ambiguity in the calculated $^7$Li abundance due to this reaction, and promises substantial improvements in the $^7$Li BBN prediction. Secondly, more accurate abundance predictions of primordial isotopes are also crucial to probe exotic physics beyond the standard model as well as to constrain cosmological parameters~\citep{pos10,fie11,coc13,kus14,coc14,yam14,hou17,luo19,mos20}. For these reasons, this work is important for the continued developments of other interdisciplinary fields such as astronomy, cosmology, and particle physics.

The current $^7$Li(d,n)2$^4$He reaction rate most widely used in BBN simulations is taken from~\citet{Boy93} (hereafter referred to as BM93). Differing from the rate compiled by~\citet{cf88} (hereafter CF88), which only considered the direct component, the BM93 rate not only updated the direct reaction rate, but also took the contributions from the 280 keV and 600 keV resonances into account. Theoretically speaking, the BM93 rate should be more reliable compared with the evaluation from CF88. Nevertheless, we found that a significant overestimation exists in their assessment, which could potentially impact BBN nucleosynthesis. In this work, the $^7$Li(d, n)2$^4$He reaction rate is investigated systematically and comprehensively, and the separate contributions from the direct components and resonances near the deuteron threshold are studied individually.
It is well known that the uncertainties of every individual component come from the uncertainties from its own input parameters. For the purpose of getting more reasonable uncertainties of the total reaction rates, a Monte Carlo approach is used to obtain the total $^7$Li(d,n)2$^4$He reaction rate and its corresponding error. In order to study the impact of the new reaction rate on the abundances of primordial nuclei, we perform  detailed BBN calculations using two types of models: a uniform density distribution model and a nonuniform density model.

The paper is organized as follows. In section 2, we introduce the basic formalism for the resonant reaction cross section and its relation to the astrophysical reaction rate. In section 3, based on an elaborate investigation of each individual term which contributes to reaction rates, we derive the cross section of the $^7$Li(d,n)2$^4$He reaction and its corresponding uncertainties, further obtaining the new astrophysical reaction rate of $^7$Li(d,n)2$^4$He. In section 4,  we perform the BBN simulations with a uniform and a nonuniform baryon distribution to investigate the impact of our new $^7$Li(d,n)2$^4$He reaction rate on primordial yields.  Our conclusions are summarized in the last section.

\section{Astrophysical reaction rate}
The thermonuclear rate is calculated from the reaction cross section $\sigma(E)$ by integration over the Maxwell-Boltzmann distribution of the interacting particles in a stellar environment with a temperature $T $ ~\citep{rol88,ili07}
\begin{equation}
\label{eq1}
\left\langle\sigma v\right\rangle=\sqrt{\frac{8}{\pi\mu(kT)^3}}\int_{0}^{\infty}\sigma(E)E\mathrm{exp}\left(-\frac{E}{kT}\right)\,dE,
\end{equation}
where $\mu$ is the reduced mass, $N_A$ is Avogadro's number, and $k$ is the Boltzmann constant. Obviously, the reaction cross section $\sigma(E)$ and its energy dependence are the key parameters for determining the reaction rates.  $\sigma(E)$ is comprised of resonant and direct reaction cross section.

\subsection{Resonant cross section}
Differing from a one-step process without the formation of an intermediate compound nucleus (direct reaction), a resonant reaction proceeds through the formation of a compound nucleus in the entrance channel which subsequently decays to the exit channel. The resonant cross section is described by a Breit-Wigner single level formula
\begin{equation}
\label{eq1000}
\sigma =\frac{\pi}{2\mu E}\frac{\omega\Gamma_{in}\Gamma_{out}}{(E-E_r)^2+\Gamma^2_{tot}/4},
\end{equation}
where the first term is the upper limit for the cross section (i.e., the geometrical cross section), $E$ is the energy in the center of mass frame ($CM$), $E_r$ is the resonance energy, and $\Gamma_{in}$ and $\Gamma_{out}$ are the widths of the entrance channel and exit channel, respectively. The total resonance width of the state is defined as $\Gamma_{tot}$ = $\Gamma_{in}+\Gamma_{out}+ \cdots$. The statistic factor $\omega$ is defined as
\begin{equation}
\label{eq1001}
\omega =\frac{(2J_C+1)(1+\delta_{01})}{(2J_0+1)(2J_1+1)},
\end{equation}
which takes into account the angular momenta $J_0$ and $J_1$ of the colliding nuclei and the angular momentum $J_C$ of the excited state in the compound nucleus. The factor $(1+\delta_{01})$ is included since the cross section for identical particles in the entrance channel increases by two times.

The resonance width $\Gamma_{in}$($\Gamma_{out}$) can be parameterized by the dimensionless reduced width $\theta^2$, which incorporates all the unknown properties of the nuclear interior,
\begin{equation}
\label{eq1002}
\Gamma_{in} =\frac{3\hbar v}{R}P_l(E,R)\theta^2_{in},
\end{equation}
where $v$ is the relative velocity in the $CM$ frame and $R$ is the interaction radius. The function $P_l(E,R)$ refers to the Coulomb and centrifugal barrier penetrability given by
\begin{equation}
\label{eq1003}
P_l(E,R) =\frac{1}{[F_l(E;R)^2+G_l(E;R)^2]}.
\end{equation}
Here, $F_l(E;R)$ and $G_l(E;R)$ are the regular and irregular Coulomb wave functions, respectively.

\subsection{Resonant reaction rate}

\subsubsection{Narrow resonant reaction rate}
Broadly speaking, the astrophysical reaction rate should be obtained by performing strict numerical integration following Eq.~1. However, in the case of a narrow resonance for which the width of the resonance is much smaller than resonance energy,  the expression for the rate can be rewritten as

\begin{equation}
\label{eq1}
\left\langle\sigma v\right\rangle=\sqrt{\frac{2\pi}{(\mu kT)^3}}{exp}\left(-\frac{E_r}{kT}\right)\omega\frac{\Gamma_{in}\Gamma_{out}}{\Gamma_{tot}}
2\int_{0}^{\infty}\frac{\Gamma_{tot}/2}{(E_r-E)^2+\Gamma^2_{tot}/4}\mathrm\,dE,
\end{equation}
since the partial width and the energy factor from the MB distribution are approximately constant over the whole total width of the resonance.

By introducing the concept of resonance strength with definition of $\omega\gamma=\omega\Gamma_{in}\Gamma_{out}/\Gamma_{tot}$ and the integral in~Eq. 6 can be calculated analytically, the narrow resonance reaction rate can be simplified as

\begin{equation}
\label{eq1}
\left\langle\sigma v\right\rangle=\left(\frac{2\pi}{\mu kT}\right)^{3/2}\hbar^2 \omega\gamma {exp}\left(-\frac{E_r}{kT}\right).
\end{equation}

\subsubsection{Reaction rate of broad resonance and sub-threshold resonance}

For the case of a broad resonance where the resonance width is not much smaller than the width of the Gamow peak for a given temperature, it can no longer be assumed that partial widths and the M-B distribution factor can be pulled out front from the integration as constants. In such a case, the reaction rates must be calculated by numerical integration using Eq.~1. Similar cases occur to subthreshold resonances where the compound level lies below the particle threshold and the reaction can proceed via the high energy wing of the resonance extending over the particle threshold. Likewise, the energy dependence of the partial and total widths is required as well. Therefore, we just need to follow the same procedure used for broad resonance to calculate the contribution from a subthreshold resonance to the reaction rate.

\section{Derivation of the reaction rate for $^7$Li(d,n)2$^4$He}

It is well known that the astrophysical reaction rate is mainly determined by the reaction cross section in the energy region close to the threshold. For the reaction $^7$Li(d,n)2$^4$He, it is known that there are four near-threshold resonances: the subthreshold state at -24.9 keV and three above-threshold resonances at 0.28, 0.6, and 0.8 MeV, respectively. However, the properties of these resonances are still ambiguous and the partial widths of some resonances remain unknown up to now. In this section, we will study each of them in detail utilizing the results from recent experiments.

\subsection{ The cross section break-down}

\subsubsection{Consideration of the subthreshold resonance}

As noted in the introduction, the newest $^7$Li(d,n)2$^4$He astrophysical reaction rate widely used in BBN simulations is from the estimation of BM93, which includes the contributions from both direct components and resonances at 280 keV and 600 keV. Compared with the rates from CF88, wherein only the contribution from the direct term is considered, the BM93 rates integrate the contributions from resonances near the threshold for the first time which means, at least in principle, it should be reliable. However, neither rate includes the contribution from the $^9$Be resonance state at 16.671 MeV~\citep{til04}, which is only 24.9 keV below the energetic threshold of the $^7$Li + d reaction; this is most likely due to the inaccurate information regarding energy levels of the 9Be nucleus at that time. Fortunately, fruitful follow-up experimental studies on the energy level of $^9$Be made it possible to assess its contribution to the $^7$Li + d reaction rates.

Owing to the absence of a Coulomb barrier for neutron emission from the $^9$Be compound system, it is conventionally thought the d+$^7$Li reaction proceeds mainly through intermediate states in $^8$Be by the $^7$Li(d,n)$^8$Be($\alpha$)
$^4$He reaction sequence, and not through intermediate states in $^5$He by the $^7$Li(d,$\alpha$)$^5$He(n)$^4$He sequence. Nevertheless, completely differing from our conventional understanding, the recent experimental result from~\citet{rij19} strongly indicated that $\alpha$-decay dominates for the 16.849 MeV, 5/2$^+$ state in $^9$B (actually corresponding to the 16.71 MeV state in energy level diagram of $^9$B in NNDC), which is regarded as the mirror state of 5/2$^+$ state at 16.671 MeV in $^9$Be. According to the mirror symmetry principle, the J$^\pi$=5/2$^+$ subthreshold state in $^9$Be should primarily decay by $\alpha$-emission as well, and the counterpart $^5$He then subsequently splits into neutron and $\alpha$. Therefore, nucleosynthesis calculations should also take into account the $^7$Li(d,$\alpha$)$^5$He reaction although the final products are the same as the $^7$Li(d,n)$^8$Be reaction.

For the purpose of calculating the separate contribution from the subthreshold resonance to the $^7$Li(d,n)2$^4$He reaction rate, the energy level information of this state including spin, parity and  partial width for d, n, and $\alpha$ decay of this resonance are required. It is known from NNDC that the spin and parity of the 16.67 MeV resonant state of $^9$Be are determined as J$^\pi$=5/2$^+$, while the relevant knowledge of partial decay widths of this state remains unknown. Thus, we have to derive the widths of these particle decays using resonance theory in combination with relevant width information of its mirror state in $^9$B.

From section 2, we know that the partial decay width $\Gamma_i$ ($i=\alpha, d, n$) essentially depends on three aspects: relative velocity $v$, penetration factor $P_l(E,R)$ and reduced width $\theta^2$. Among them, relative velocity $v$ can be obtained easily and $P_l(E,R)$ can be calculated analytically for neutron emission, but for the case of charged nuclei, we have to resort to a numerical calculation instead of analytical approximation in order to obtain the $P_l(E,R)$ with relatively high accuracy. In our calculation, a code following the formalism depicted in ~\citet{ili97} is used to calculate $P_l(E,R)$. Regarding the reduced width $\theta^2$, it reflects a measure of the degree to which actual quasi-stationary state can be described by the motion of particle $a$ and the residual nucleus $X$ in a potential. In principle, it can be estimated on the basis of a nuclear potential approximated as a square well and assuming average level distance~\citep{bla10}. Nevertheless, the value derived from experiments would be of high priority for its use in calculations.

It is well known that a mirror state is referred to as an analogue state at nearly the same excitation energy in mirror nuclei pairs, which can be inter-transformed by exchanging the role of protons with neutrons. According to mirror symmetry, the properties and configuration of the mirror states in the mirror pair of $^9$Be and $^9$B should be identical apart from the Coulomb effects. Thus, it is expected that the $\theta^2$ value holds constant for identical particle decay from the mirror states pair of $^9$B at 16.849 MeV and $^9$Be at 16.67 MeV.
Therefore, the reduced width $\theta^2$, which will be used to calculate the partial width of the subthreshold level (16.67 MeV) of $^{9}\mathrm{Be}$, can be extracted directly from the relevant widths information of its mirror state in $^{9}\mathrm{B}$ (16.849 MeV, 5/2$^+$). Fortunately, the partial widths for the mirror state in $^{9}\mathrm{B}$ are available thanks to recent cross section measurements of the $^7$Be + d reaction~\citep{rij19}. In their analysis, it is shown that the (d,$\alpha$) channel dominates relative to the (d,p) channel and the values of $\Gamma_{\alpha}$ and $\Gamma_d$ are suggested to be 50 keV and 3.3 keV, respectively. $\Gamma_p$ is only 1 keV, implying the contributions from the (d,p) channel is negligible. Using Eq.~4, the reduced width $\theta_i^2$ ($i=\alpha, d, p$) values for the mirror state in $^9$B of 16.849 MeV are obtained and listed in Table~\ref{tab1}. Here, the uncertainty of $\theta_d^2$ is mainly caused by the influence of the Coulomb penetration factor $P_l$ from the 5 keV uncertainty of the resonance energy. However, this hardly makes a visible impact on $\theta_a^2$ and $\theta_p^2$ because several keV uncertainty in the energy level can be totally neglected with respect to the huge energy release for $\alpha$- and $p$-decay. In the present evaluation, we mainly consider the $^7$Li(d,$\alpha$)$^5$He channel since the neutron decay is negligible relative to $\alpha$-decay. Using the above reduced widths of $d$ and $\alpha$ from the mirror state in $^9$B, as shown in the first row of Table~\ref{tab2}, the cross section for the $^7$Li(d, $\alpha$)$^5$He reaction proceeding through the subthreshold compound nucleus $^9$Be can be obtained using Eq.~1.

\begin{table*}[]
\scriptsize
\caption{\label{tab1}The reduced width of different particle decays for the 16.849 MeV level in $^9$B.}
\begin{tabular}{lcccccccccl}
\hline
\hline
 $E_r$ (MeV) & $J^{\pi}$ & $\Gamma_{p_1}$ & $\Gamma_d$ & $\Gamma_{\alpha}$ & $\theta^2_{p_1}$ & $\theta^2_d$ & $\theta^2_{\alpha}$\\
 0.361(5)& 5/2$^+$ & 1 & 3.3 & 50 & 3.64$\times$10$^{-5}$ & 0.119$\pm$0.008 & 3.22$\times$10$^{-3}$ \\
\hline
\end{tabular}
\end{table*}

\subsubsection{The 600-keV resonance}

The resonance, 600 keV above the deuterium threshold (16.6959 MeV), corresponding to the 5/2$^-$ exited state of $^9$Be at 19.298 MeV, is thought to be very significant in the evaluation by BM93 since they thought its contribution to the $^7$Li+d reaction rates dominate within the temperature range of BBN interest. In their evaluation, the total cross section at resonance energy $E_r$=600 keV is determined to be 420 mb, which was obtained by multiplying the value of the cross section measured at 0$^\circ$ by 4$\pi$ on the assumption that the differential cross section is isotropic~\citep{sla57}.
According to the Breit-Wigner single resonance formula Eq.~2, we know that the reaction cross section $\sigma_r$ will reach a maximum when the condition of $\Gamma_{in}$=$\Gamma_{out}$=$\Gamma$/2 and $\Gamma$=$\Gamma_{in}$+$\Gamma_{out}$ is satisfied. We also know that the total width of this state is determined as 200 keV from~\citet{til04}, so if one sets $\Gamma_{in}$=$\Gamma_{out}$=100 keV, the obtained maximum limit of the cross section at $E_r$=600 keV should be 357 mb, which is still smaller than the value of 420 mb adopted in BM93. If all of these considerations are correct, the cross section for the resonance at 600 keV is overestimated in the evaluation of BM93.  The reason for this overestimation is not obvious. A possible reason could be attributed to the assumption of isotropy for the angular distribution.

We now use an indirect method to reevaluate the contribution from the 600 keV resonance on the cross section of $^7$Li(d,n). In contrast to the situation of the $^7$Li(d,n) cross section at $E_r$=600 keV which has a lack of sufficient experimental data, the cross section for $^7$Li(d,p) from this resonance has been measured extensively. The currently existing values for the measured $^7$Li(d,p) cross section range from a maximum value of 211$\pm$15 mb to a minimum of 110$\pm$22 mb~\citep{ade98}, and the cross section for this resonance recommended by~\citet{ade98} is 147$\pm$11 mb based on the comprehensive consideration of previous measurements. Another new cross section measurement of this 600 keV resonance, which is free from the effect of backscattering, presented a slightly bigger value of 155$\pm$8 mb~\citep{wei98}. Here, the two above proposed cross sections lead to an average value of $\sigma_{d,p}=151\pm10$ mb, which will be used in our next calculation.

The recent measurements of resonances in $^9$Be around the proton threshold present the specific partial widths of (p, d), (p, $\alpha$) and (p, p) of the 19.298 MeV excited state ($E_r$=600 keV for deutron threshold) via multichannel R-matrix analysis of the experimental data~\citep{lei18}. The final values of $\Gamma_{p}$, $\Gamma_{d}$, $\Gamma_{\alpha}$ for this state and the associated uncertainties are determined to be 40$\pm$10, 150$\pm$7, and 20$\pm$3 in keV, respectively. Using the values of $\Gamma_{d}$ and $\Gamma_{p}$ given above, we attempted to reproduce the value of 151$\pm$10 mb by adjusting the values of $\Gamma_{d}$ and $\Gamma_{p}$ within their own uncertainties and finally $\Gamma_{d}$= 143 keV and $\Gamma_{p}$ = 30 keV are proposed.

Recalling the specific derivation of the resonant cross section of $^7$Li(d,n) from the 17.298 MeV resonance in the BM93 estimation, one point we need to highlight is that $\Gamma_{\alpha}$ was thought to be negligible compared to $\Gamma_n$. However, we found there is no definite evidence to support their conclusion in the literature~\citep[e.g.,][]{Heg73}. On the contrary, the experimental results from ~\citet{Heg73} show that the alpha emission accounts for a significant fraction relative to neutron emission. Our conclusions are not affected by which channel dominates between $^7$Li(d,n) and $^7$Li(d,$\alpha$) since the final reaction products in a three-body form will be identical for both the two channels. The sum of the contributions from these two channels can be taken as the resonant cross section for $^7$Li(d,n) or $^7$Li(d,$\alpha$) at 17.298 MeV. As indicated in ~\citet{til04}, the $\gamma$ decay width is only at the level of several eV so the total width ($\Gamma$=200 keV) of the 600 keV resonance consists almost entirely of $\Gamma_{p}$, $\Gamma_{n}$, $\Gamma_{d}$, $\Gamma_{\alpha}$. Note that the partial widths $\Gamma_{p}$, $\Gamma_{d}$ have been set as introduced above, and all of the remaining fraction of the total width $\Gamma$ (subtracting $\Gamma_{p}$ and $\Gamma_{d}$) can be taken as the value of $\Gamma_{n}$ or $\Gamma_{\alpha}$. For this reason, we can set the the upper limit of $\Gamma_{\alpha}$ as $\Gamma$ - ($\Gamma_{d}$ + $\Gamma_{p}$). Then the reduced widths $\theta_i^2$ ($i = p, d$) and the upper limit of $\theta_{\alpha}^2$ can be obtained via the partial width formula Eq.~4, as listed in the second row of Table~\ref{tab2}.

Unlike the reaction rate from a single narrow resonance, which can be directly obtained via ~Eq.7,
the calculation of the reaction rate for the 17.298 MeV resonance relies on~Eq.1  since this resonance is confirmed to be broad. Similar to the case of subthreshold resonance, the knowledge of the dimensionless reduced widths $\theta^2$ for different decay channels are required. Using our deduced values of $\Gamma_{d}$ and $\Gamma_{p}$ for the resonance at $E_r$=600 keV in combination with the partial width formula(~Eq.4), the reduced widths $\theta_i^2$ ($i = p, d$) can be obtained, as listed in Table~\ref{tab2}. If we take the remaining width $\Gamma$-($\Gamma_{d}$ + $\Gamma_{p}$) as the the upper limit of $\Gamma_{\alpha}$ or $\Gamma_{n}$, we can then obtain the maximum value of the cross section for the $^7$Li(d,n)2$^4$He from the 600 keV resonance.

\begin{table*}[]
\caption{\label{tab2}Resonance properties (energies in MeV, widths in keV) considered in the present calculation. Energies are given with respect to the $^7$Li+D threshold. Note: the value with an $^{\star}$ denotes the upper limit for the given quantity and specific particle decay.}

\begin{tabular}{lcccccccccl}
\toprule\toprule
 $E_r$ (MeV) & $J^{\pi}$ & $\Gamma_p$  & $\Gamma_d$ & $\Gamma_{\alpha}$ & $\theta^2_p$ & $\theta^2_d$ & $\theta^2_{\alpha}$\\
\hline
-0.0249$\pm$0.008 & 5/2$^+$   &            &        &            &      & 0.119$\pm$0.008      & 3.22$\times$10$^{-3}$ \\
0.6$\pm$0.005    & 5/2$^-$   & 30         & 143    & 27$^{\star}$         & 0.186$\pm$0.006     & 0.099$\pm$0.0014       & 1.872$\times$10$^{-3}$$^{\star}$ \\
 0.8$\pm$0.005    & 7/2$^+$   & 1$\pm$0.2  &7$\pm$3 & 39$\pm$4   &            &            &                \\
\bottomrule
\end{tabular}
\end{table*}

\subsubsection{The 800 keV resonance}
The 800 keV resonance corresponds to the excited state of $^9$Be at 17.493 MeV, with recommended J$^{\pi}=7/2^+$ assignment in~\citet{til04}. The nature of this state, like partial decay widths, is still not well understood, despite the fact the total width is confirmed to be 47 keV. Fortunately, information regarding the partial width of this state is given by~\citet{lei18} and will be used directly in our evaluation, as shown in third row of Table~\ref{tab2}.
We emphasize again that the contribution of $^7$Li(d,$\alpha$)$^5$He is regarded as equivalent to the $^7$Li(d,n)2$^4$He channel since the final reaction products of both reactions are identical. Through a similar process to that in the previous section,  the cross section of the 17.493 MeV resonance can be obtained.

\subsubsection{The direct contribution}
For the sake of comparison with previous results and convenience of discussion below, we here write the cross section in the form of $\sigma(E) = S(E)E^{-1} e^{-2\pi\eta}\sigma(E)$, where $S(E)$ is the astrophysical S-factor and $\eta$ is the Sommerfield parameter.
Examining the previous evaluations of direct contributions from $^7$Li(d,n) in literature~\citep{cf88, Boy93}, it is clear that the source data used to determine the direct S-factor in BM93 originates from ~\citet{sla57}, which is exactly the same one used to derive the cross section of the 600 keV resonance, while that of CF88 remains unclear. We investigate the relevant literature and assume that the data originates from~\citet{bag52}. This assumption is also confirmed by comparing the CF88 rate to the results from a numerical integration over the cross section data in low energy regions of~\citet{bag52}. The direct S-factor $S(0)$ determined in CF88 is about 33.9 MeV-barn, which is about 2 times that derived in~\citet{Boy93} based on data between 1.6 and 2.0 MeV of deuteron energy. It is difficult to conclude which is correct since both of them have their own intrinsic drawbacks.

The uncertainties of the CF88 direct S-factor mainly stem from the constraints of measurements at very limited solid angle (90$\pm$20 degree) and the assumption of isotropic angular distribution, which have proven to result in S-factor overestimation(see section 3.1.2 on the 600 keV resonance). In addition, at low energies, extra contributions from possible resonances near the threshold, such as yet to be identified subthreshold resonances and effects from electron screening, will both lead to overestimation of the S-factor from direct term. Differing from the case of CF88, the S-factor of direct term from~\citet{Boy93} is determined as 17 MeV-b based on the fact that the derived S-factor values almost appear to be a constant in the energy range from 1.6 to 2.0 MeV. Theoretically speaking, this value might be more reliable compared with that from CF88 since the interference from subthreshold resonances and electron screening can be excluded. However, it is likely still overestimated. The reasons are threefold: Firstly, this value of 17 MeV-b from BM93 actually refers to the sum of the $^7$Li(d,n) and $^7$Li(d,p) channels, while the actual contribution from $^7$Li(d,n) is only about 9 MeV-b. The remaining part accounts for the endoergic  $^7$Li(d,p)$^8$Li reaction which does not proceed efficiently in BBN as a result of the dominance of its reverse reaction. Therefore, they should be separately treated as two different reactions whenever performing BBN network calculations, but this is not mentioned in all of the previous BBN simulations where the BM93 rate is used directly~\citep{ser04,Pis08,coc12,Arb12,Con18,pit18}. Secondly, it is well known that the cross section value from experimental measurements for a fixed reaction can reflect only the total contributions from direct components and resonant components, not the separate contribution. So the value from BM93 still includes the resonant contribution in or near the energy zone of 1.6 to 2.0 MeV. Thirdly, the same as for the case of the 600 keV resonance where the isotropic angular distribution is assumed for the same source data of cross section, it is inevitable to produce overestimation of direct S-factor. In other words, this S-factor of about 9 MeV-b is still larger than the true value taking into account only the direct contribution.

In the present work, instead of adopting the data from~\citet{bag52} and~\citet{sla57}, we choose data from the recent work where the direct S-factor is determined as S(E)=5400($\pm$1500) - 37($\pm$21)E keV$\cdot$b~\citep{sab06}. In their experiment, the cross section of $^7$Li(d,n) was measured for energies below 70 keV and the emitted neutrons were detected at eight different angles from 0$^\circ$ to 150$^\circ$, so the obtained S-factor value could exclude the influence from the assumption of the isotropic angular distribution in~\citet{bag52}. Another reason we choose this value for $S(E)$ is due to its consistency with the value estimated by using the extracted factor of overestimation to scale the $S(0)$ of the $^7$Li(d,n) channel in~\citet{bag52}. Specifically, it can be seen that both of the cross section of $^7$Li(d,p) and $^7$Li(d,n) were measured in~\citet{bag52}. The cross section of $^7$Li(d,p) at the peak of 600 keV resonance is up to 230 mb in their results, but the proceeding measurements of the $^7$Li(d,p) cross section for this resonance support the value of 147 mb recommended in~\citet{ade98}, which is about 64 percent of the value from~\citet{bag52}. Then the factor of overestimation extracted out from these two sets of $^7$Li(d,p) cross section data can be used to scale the direct S-factor of $^7$Li(d,n) derived from the low energy cross section in~\citet{bag52}. The obtained value is approximately 5700 keV$\cdot$b, basically in accordance with the value from~\citet{sab06}.

\subsection{Astrophysical S-factor and reaction rate of $^7$Li(d,n)2$^4$He}

\begin{figure}[tbp]
\begin{center}
\includegraphics[width=8.6cm]{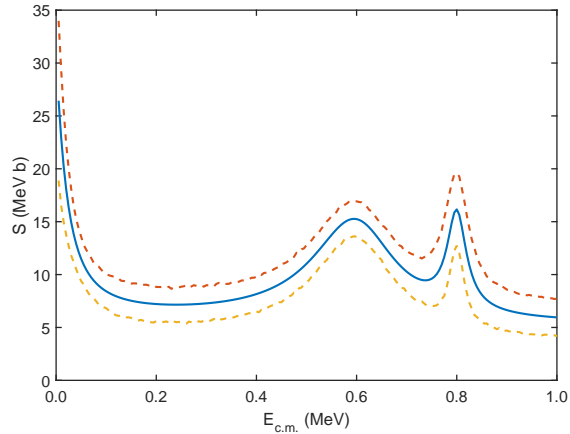}
\vspace{-1mm}
\caption{\label{fig1} The obtained astrophysical S-factor of $^7$Li(d,n)2$^4$He reaction as a function of energy corresponds to the solid blue line, while dashed red and orange lines signify the upper limits and lower limits, respectively.}
\end{center}
\end{figure}

Using the information of $\Gamma_i$ and $\theta^2_i$ listed in table~\ref{tab2} and the newly determined direct S-factor from previous section, the total $S(E)$ (or cross section $\sigma(E)$) of $^7$Li(d,n)2$^4$He and its corresponding uncertainties can be obtained via a Monte Carlo simulation considering all the the uncertainties of the resonance parameters, as plotted in Figure~\ref{fig1}. The solid blue line refers to the total S-factor for $^7$Li(d,n)2$^4$He, and the dashed red and orange lines correspond to upper and lower limits, respectively. Figure.1 shows that the rapid decrease of the S-factor in the low energy region is due to the -24.9 keV subthreshold resonance. The two peaks are the 5/2$^-$ resonances at $E_x$=17.298 MeV ($E_r$=600 keV) and the 7/2$^-$ resonance at $E_x$=17.493 MeV ($E_r$=800 keV), respectively. Here, we did not consider the 1/2$^-$ resonance at $E_x$=16.9752 MeV ($E_r$=280 keV) since its width is too narrow to make a noticeable contribution to the final results.

\begin{figure}[tbp]
\begin{center}
\includegraphics[width=8.6cm]{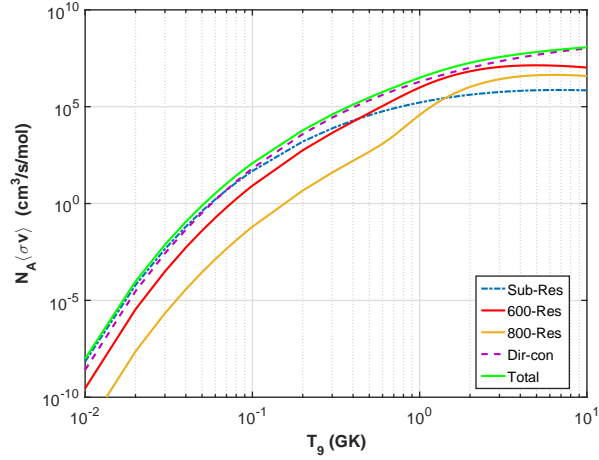}
\vspace{-1mm}
\caption{\label{fig2} The new reaction rate of $^7$Li(d,n)2$^4$He as a function of temperature in units of GK is shown by the green solid line. The rate contributions from direct reaction and various resonances are displayed separately. The dashed-dotted line shows the subthreshold resonance ($E_r$ = -25 keV). The solid red and orange lines correspond to the resonances at $E_r$ = 600 keV and 800 keV, respectively. The dashed purple line indicates the direct component.}
\end{center}
\end{figure}

\begin{figure}[tbp]
\begin{center}
\includegraphics[width=8.6cm]{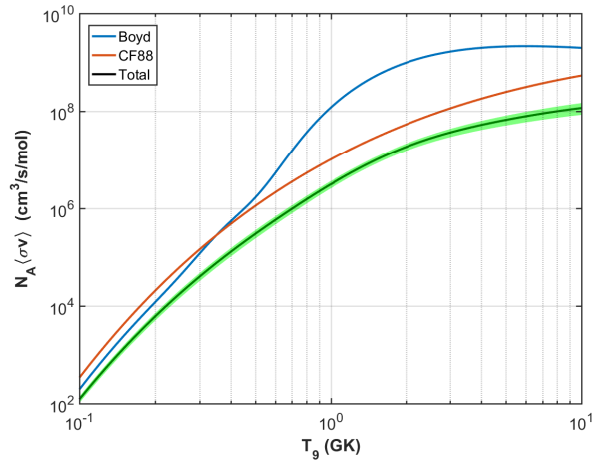}
\vspace{-1mm}
\caption{\label{fig3} Total reaction rate of $^7$Li(d,n)2$^4$He as a function of temperature in units of GK where the green shaded band is its associated uncertainties. For comparison, we also plot the previous results from CF88 and BM93.}

\end{center}
\end{figure}

The S-factor obtained above is then inserted into the reaction rate expression  to calculate the total reaction rate of $^7$Li(d,n)2$^4$He as shown by the green solid line in Figure~\ref{fig2}. We also break down our newly obtained total reaction rate in terms of the separate contributions from direct and various resonances. It is seen clearly in Figure~\ref{fig2} that the reaction rate is dominated by the nonresonant contributions from direct reaction and subthreshold resonances, rather than by the 600 keV resonance asserted in the previous evaluation from BM93. In particular, the contribution from the -24.9 keV subthreshold resonance dominates for temperatures lower than 0.07 GK. For temperatures T $\textgreater$ 0.07 GK, it is mainly contributed by direct reaction. Around the temperature of 1.6 GK, the 600 keV (5/2$^-$) broad resonance contributions is comparable with those from direct components.  The contribution from the 800 keV resonance can basically be neglected. Here, the narrow resonance at $E_r$=280 keV is neglected in our calculation since its contribution is even smaller than that from the 800 keV resonance.

The uncertainty of the S-factor shown in Figure~\ref{fig1} is also used to calculate the uncertainty for the $^7$Li(d, n)2$^4$He reaction rates, as shown in Figure~\ref{fig3} by the green shaded band. For convenience of comparison, the old rates from previous works are also included in Figure~\ref{fig3}, where the blue solid line is for the BM93 rate and the brown line is for CF88 rate. It can  clearly be seen from Figure~\ref{fig3} that our new rates, including its upper and lower limits, is overall smaller than the previous two evaluations. In particular, for temperatures in the range of BBN importance (up to about 1 GK), our new rate is about 60 times smaller than the rate widely used in current BBN simulations given by~\citet{Boy93}, which strongly motivates us to explore its impact on the production of $^7$Li during BBN.

For the sake of convenience for its use by others, the present rate can be well fitted (less than 0.11\% error in 0.01---10 GK) by the following analytic expression in the standard seven parameter format of REACLIB:

$N_A\left\langle\sigma v\right\rangle=\mathrm{exp}(45.3213+0.180629T_9^{-1}-16.8231T_9^{-1/3}
-14.9337T_9^{1/3}+1.22317T_9-0.0685717T_9^{5/3}+1.80904ln(T_9)+
         \mathrm{exp}(0.410313+0.0129319T_9^{-1}-14.3153T_9^{-1/3}+36.1545T_9^{1/3}-9.83075T_9
         -0.445434T_9^{5/3}-8.38412ln(T_9))$

\section{BBN calculation}
Using the new $^7$Li(d,n)2$^4$He reaction rate, we perform a BBN simulation to investigate its effect on the primordial $^7$Li abundance by using a modified Wagoner code with updated reaction rates. In our calculation, we choose the most up-to-date baryon-to-photon ratio $\eta_{10}$=6.104($\eta_{10}=\eta\times10^{10}$)~\citep{planc} and the newest world average for the neutron lifetime ($\tau$=879.4$\pm$0.6 s) from ~\citet{fie20}. The predicted light element abundances are shown in Table~\ref{tab3}. It is found that the adoption of the new rate hardly results in any remarkable change in the final $^7$Li yield. In particular, the abundance of $^7$Li only incrases by about 0.002\%. The present nuclear uncertainties associated with the $^7$Li+d channel have no impact on $^7$Li nucleosynthesis. The reason for such a low effect is that the direct production of $^7$Li dominates at low baryon abundances ($\eta_{10}\le3$), whereas the direct production of $^7$Be dominates at higher baryon abundances ($\eta_{10}\ge3$), as is remarked in~\citet{ste07}. In other words, the alteration of the direct production of $^7$Li caused by the new $^7$Li(d,n)2$^4$He rate can be completely neglected in regions of high baryon density.

\begin{table*}
\scriptsize
\caption{\label{tab3} The predicted abundances of primordial light elements from the standard BBN model for old and our new $^7$Li(d,n)2$^4$He reaction rate.}
\begin{tabular}{lllllll}
\hline \hline
   & D/H & $^{3}$He/H & $^{4}$He & $^{6}$Li/H & $^{7}$Li/H & $^{7}$Be/H  \\
\hline
Old & 2.567$\times10^{-5}$ & 1.019$\times10^{-5}$ & 0.247 & 1.114$\times10^{-14}$ & 3.119$\times10^{-11}$ &4.539$\times10^{-10}$ \\
New & 2.567$\times10^{-5}$ & 1.019$\times10^{-5}$ & 0.247 & 1.114$\times10^{-14}$ & 3.120$\times10^{-11}$ &4.539$\times10^{-10}$ \\
\hline \hline
\end{tabular}
\end{table*}

With this in mind, it is worthwhile to consider alternative BBN models which can produce low baryon density regions~\citep{Rau94,ori97}. One appropriate candidate where we would expect the $^7$Li+d reaction to have an impact is that of an inhomogeneous density distribution at the time of Big Bang nucleosynthesis (IBBN), in which $^7$Li(d,n)2$^4$He probably plays a pronounced role.
We adopt the same IBBN model as in~\citet{ori97}, where the model is parameterized by the density contrast between the high and low-density regions R; the volume fraction of high-density region $f_v$; the distance scale of the inhomogeneity $r$, in addition to the average baryon-to-photon ratio $\eta$ and fluctuation geometry. The present calculation is performed in a cylindrical shell geometry, where the same set of model parameters ($R, f_v, r$) as in ref ~\citep{ori97} are used to characterize the density nonuniformity, where $R$=10$^6$, $f_v$= 0.15 and $r$=10$^6$ meters, respectively, as an illustrative example.  These parameter values are chosen such that the observed constraints on light elemental abundances, except for A = 7, are maximally satisfied. However, for the parameters $\eta$, $\tau_{n}$ and $N_{\nu}$, the same values are adopted as those used in HBBN. The fluctuations are divided into 16 zones of variable width as described in~\citet{mat90}, where the baryon density of the zone ($z_{i}$) increases with the zone number $i$ from 1 to 16. The relevant thermonuclear reaction rates are the same as in the homogeneous BBN simulation.

\begin{table*}
\scriptsize
\caption{\label{tab4} The predicted abundances of the primordial light elements for our inhomogeneous BBN model with $R$=10$^6$, $f_v$= 0.15 and $r$=10$^6$ meters. $Y_o$ refers to the abundance of the specific nuclide obtained using the $^7$Li(d,n)2$^4$He rate from BM93, while $Y_n$ is the nuclide's abundance using our new $^7$Li(d,n)2$^4$He rate.  m denotes the exponent in the power of 10.}
\begin{tabular}{ccccccccccccccc}
\hline \hline
 Nuclide &  \multicolumn{4}{c}{zone1} &   &\multicolumn{4}{c}{zone16} & &\multicolumn{4}{c}{Average}\\
\cline{2-5}  \cline{7-10} \cline{12-15}
  & $Y_{o}$ & $Y_{n}$  &$m$ & $\frac{Y_n-Y_o}{Y_o}$ && $Y_{o}$ & $Y_{n}$ &$m$& $\frac{Y_n-Y_o}{Y_o}$ && $Y_{o}$ & $Y_{n}$ &$m$& $\frac{Y_n-Y_o}{Y_o}$\\
\hline

$^7$Li/H($\times$10$^{-m}$)  & 1.82   & 2.82 &9  &55\%  && 3.95  & 3.96 &11 &0.25\% && 1.26 &1.70 &10 &35\%   \\
$^7$Be/H($\times$10$^{-m}$)  & 9.38   & 9.38 &12 &0\%  && 7.38  & 7.38 &10 &0.0\% && 2.77 &2.77 &10 &0\%   \\
$^9$Be/H($\times$10$^{-m}$)  & 1.46   & 2.28 &14 &56\%  && 5.30  & 5.40 &18 &1.9\% && 0.71 &1.02 &15 &44\%   \\
$^{10}$B/H($\times$10$^{-m}$)& 2.59   & 4.09 &17 &58\%  && 4.46  & 4.52 &21 &1.3\% && 1.34 &1.94 &18 &45\%   \\
$^{11}$X/H($\times$10$^{-m}$)& 1.36   & 2.13 &16 &57\%  && 4.09  & 4.09 &17 &0.0\% && 2.67 &2.96 &17 &11\%   \\
$^{12m}$X/H($\times$10$^{-m}$)& 4.84   & 13.0 &16 &168\%  && 3.48  & 4.34 &15 &24\% && 6.89 &9.38 &15 &36\%   \\

\hline \hline
\end{tabular}
\end{table*}

Firstly, we investigate how it affects $^7$Li production in local regions, and two extreme cases are chosen: a high-density zone and a low-density zone.
Table~\ref{tab4} shows the predicted abundances of primordial nuclides for the low-density region in the columns labeled zone1 and those for the high-density region (zone16), respectively. We only show the results for $^7$Li, $^7$Be, $^9$Be, $^{10}$B, nuclides A=11 and A$\geq$12 (marked as 12m in Table~\ref{tab4}) since our new $^7$Li(d,n)2$^4$He rate has no impact on the production of other primordial isotopes. It can be seen from zone1 in Table~\ref{tab4} that the adoption of our new $^7$Li(d,n)2$^4$He rate increases the abundances of $^7$Li, $^9$Be, $^{10}$B, nuclides A=11 and A$\geq$12 by about a factor of 2, compared to the abundances obtained using the old rate. The reason for the growth of $^7$Li can be attributed to a smaller $^7$Li(d,n)2$^4$He rate with a weaker capability of $^7$Li destruction, resulting in bigger $^7$Li production. Likewise, more $^7$Li will regulate the reactions flows moving towards the direction of synthesizing more nuclei with A$\textgreater$7. This explains why the abundances of almost all of the light nuclides heavier than $^7$Li are increased. In comparison to the other extreme case, these features exhibited in the low-density zone will disappear in the high-density zone (zone16), as shown clearly in Table~\ref{tab4}. This is owing to the fact that the production of $^7$Be dominates that of $^7$Li in the high-density regions, and therefore the production of A$\textgreater$7 nuclei is mainly through $^7$Be involved reactions rather than $^7$Li reactions. Thus, the impact from the variation of the $^7$Li(d,n)2$^4$He rate can be neglected totally in a high-density region. In order to assess its net impact on the final yields of primordial nuclei, a weighted average of abundances is calculated over the 16 zones for the entire fluctuation region, shown in the columns labeled 'Average' in Table~\ref{tab4}. The results show that the final $^7$Li abundance ($^{7}$Li/H+$^{7}$Be/H) increases by 10\% and the abundances of light nuclides with A$\textgreater$7 also increases by about 40\% if we assume an inhomogeneous density distribution during the epoch of BBN.

\section{Conclusion}
Starting from recent experimental measurements on low energy excited states of the mirror nuclei pair $^9$Be and $^9$B, we make a comprehensive analysis of the nature of near deuteron-threshold resonant states of $^9$Be. For this reason, it is important to reevaluate the $^7$Li(d,n)2$^4$He reaction rate since it plays a pivotal role in the destruction of $^7$Li during BBN.  For the first time, we present the experimentally constrained uncertainties associated with this important reaction rate. It is shown that both our newly obtained reaction rate and corresponding uncertainties show a remarkable departure from earlier evaluations.
In particular, our rate is a factor of 60 times smaller than the most widely used rate (BM93) in current BBN simulations. The cross section break-down shows that the subthreshold resonance omitted from previous evaluations dominates in the temperature range below T$_9$=0.07, while the rate for T$_9$\textgreater0.07 is mainly determined by direct reaction, not by the 600 keV resonance previously thought in BM93. In order to figure out the impact of this new reaction rate, we perform simulations using two different BBN models: a uniform density model and a  non-uniform density model. The results obtained can be summarized by the following points: the adoption of the new $^7$Li(d,n)2$^4$He rate increases by 0.002\% in the final $^7$Li yield for Standard BBN models. However, for models of inhomogeneous density distribution, it can lead to about a 10\% increase in $^7$Li production and a 40\% increase in the final primordial abundances of light nuclei with A$\textgreater$7 compared to calculations using old reaction rates. Such an increase is due to the impact in the low-density zones, where the $^7$Li yield increases by a factor of 1.55. Therefore, our results confirm the existence of the cosmological lithium problem.

\acknowledgments
We are grateful to R.N. Boyd for helpful discussions and advice. We also thank C.X. Yuan for his shell model calculation on properties of $^9$Be excited states. This work was financially supported by the Strategic Priority Research Program of Chinese Academy of Sciences Grant No. XDB34020204, and the Youth Innovation Promotion Association of Chinese Academy of Sciences under Grant No. 2019406, and the National Natural Science Foundation of China under Grant Nos. 11705244, 11490562 and 11961141004, and in part by the National Science Foundation under Grant No. OISE-1927130 (IReNA) and Grants-in-Aid for Scientific Research of The Japan Society for the Promotion of Science (20K03958, 17K05459). M.P. thanks the support to NuGrid from STFC (through the University of Hull's Consolidated Grant ST/R000840/1), and access to {\sc viper}, the University of Hull High Performance Computing Facility. M.P. acknowledges the support from the "Lendulet-2014" Program of the Hungarian Academy of Sciences (Hungary), from the ERC Consolidator Grant (Hungary) funding scheme (Project RADIOSTAR, G.A. n. 724560), by the National Science Foundation (NSF, USA) under grant No. PHY-1430152 (JINA Center for the Evolution of the Elements). M.P. also thanks the UK network BRIDGCE and the ChETEC COST Action (CA16117), supported by COST (European Cooperation in Science and Technology). C.A.B acknowledges support from the U.S. DOE Grant No. DE-FG02-08ER41533 and funding contributed by the LANL Collaborative Research Program by the Texas A\&M System National Laboratory Office. Y.L. thanks the support from JSPS KAKENHI Grant No. 19J22167.

\end{document}